\documentclass[mathleft]{an}

\usepackage{graphicx}
\newcommand{\teff}{$T_{\rm eff}$}
\newcommand{\logg}{$\log g$}

\newcommand{\msun}{$M_{\odot}$}

\begin{document}

% The following seven commands are intended for editorial usage and should be 
% ignored by the author(s).

\Pagespan{000}{}% Document's page range. 

% If second parameter is left empty, the last page is computed automatically.

\Yearpublication{2010}%
\Yearsubmission{2010}%
\Month{11}%   
\Volume{...}%  
\Issue{..}%

% \DOI{This.is/not.aDOI}% 

\title{V391~Peg: identification of the two main pulsation modes from
ULTRACAM {\it u'g'r'} amplitudes\thanks{Based on observations obtained at the 
4.2 m WHT}}

\author{R. 
Silvotti\inst{1}\fnmsep\thanks{\email{silvotti@oato.inaf.it}\newline}
%Example 
%for footnote, note the usage of the \texttt{fnmsep}
%command as separator between institute number and footnote mark} 
\and  S. K. Randall\inst{2}
\and  V. S. Dhillon\inst{3}
\and  T. R. Marsh\inst{4}
\and  C. D. Savoury\inst{3}
\and  S. Schuh\inst{5,6}
\and  G. Fontaine\inst{7}
\and  P. Brassard\inst{7}
%\and  S. P. Littlefair\inst{3}
}
\titlerunning{V391~Peg: identification of the two main pulsation modes from
ULTRACAM data}
\authorrunning{R. Silvotti et al.}
\institute{
INAF--Osservatorio Astronomico di Torino, strada dell'Osservatorio 20, 
10025 Pino Torinese, Italy
\and
European Southern Observatory, Karl-Schwarzschild-Str. 2, 85748 Garching bei 
M\"unchen, Germany 
\and 
Department of Physics and Astronomy, University of Sheffield, Sheffield S3 
7RH, UK
\and 
Department of Physics, University of Warwick, Coventry CV4 7AL, UK
\and
Georg-August-Universit\"at G\"ottingen, Institut f\"ur Astrophysik,  
Friedrich-Hund-Platz 1, 37077 G\"ottingen, Germany 
\and
Institut f\"ur Astronomie und Astrophysik, Kepler Center for Astro and 
Particle Physics, Eberhard-Karls-Universit\"at, Sand 1, 72076 T\"ubingen, 
Germany 
\and
D\'epartement de Physique, Universit\'e de Montr\'eal, C.P. 6128, 
Succ. Centre-Ville, Montr\'eal, Qu\'ebec H3C 3J7, Canada
}

\received{16 Apr 2010}
\accepted{}
\publonline{later}


\abstract{%
V391~Peg (HS~2201+2610) is an extreme horizontal branch subdwarf B (sdB) star,
it is an hybrid pulsator showing $p$- and $g$-mode oscillations, and hosts
a 3.2/sin$i$ M$_{Jup}$ planet at an orbital distance of about 1.7 AU.
In order to improve the characterization of the star, we measured the
pulsation amplitudes in the {\it u'g'r'} SLOAN photometric bands using 
ULTRACAM at the William Herschel 4.2 m telescope and we compared them with
theoretical values.
The preliminary results presented in this article conclusively show that the 
two main pulsation periods at 349.5 and 354.1~s are a radial and a dipole mode
respectively. This is the first time that the degree index of multiple modes 
has been uniquely identified for an sdB star as faint as V391~Peg (B=14.4), 
proving that multicolor photometry is definitely an efficient technique to 
constrain mode identification, provided that the data have a high enough 
quality.}

\maketitle

\section{Introduction}

About half of field sdB stars, which reside in binary systems, can form 
through common envelope ejection or stable Roche lobe overflow (Han et al. 
2002, 2003).
It is more difficult to explain the formation of a single sdB star.
Two scenarios are possible: the merger of two low-mass helium white dwarfs and
an early hot helium flash; but both are not fully consistent with the 
observations.
A recent review on these arguments is given by Heber (2009).
Another possibility, suggested by Soker (1998), is that the huge mass loss 
needed to form an sdB star is triggered by low-mass bodies, planets or brown 
dwarfs (BDs).
Although this possibility has not been tested by detailed models yet,
a planet to the pulsating sdB star V391~Peg (Silvotti et al. 2007) and three
circumbinary planets/BDs to the eclipsing sdB+M systems HW~Vir (Lee et al. 
2009) and HS~0705+6700 (Qian et al. 2009) suggest that sdB planets/BDs could 
be a relatively common phenomenon (see also Bear \& Soker 2010).
A systematic search for sdB substellar objects around 4 sdB stars using the 
timing method is the main  goal of the EXOTIME project (Schuh et al. 
2010, Benatti et al. 2010).

\vspace{1mm}

V391~Peg (HS~2201+2610) is a particularly interesting system formed by an sdB
star and a 3.2/sin$i$ M$_{Jup}$ planet orbiting the host star in 3.2 years
at a distance of about 1.7 AU (Silvotti et al. 2007).
The sdB star is a hybrid pulsator showing $p$ and $g$-mode oscillations at 
the same time ({\O}stensen et al. 2001, Lutz et al. 2009), offering a unique 
opportunity to characterize the host star through asteroseismic methods.
A preliminary mode identification of the higher pulsation frequencies 
($p$-modes) was proposed in Silvotti et al. (2002): the two main pulsation 
periods of 349.5 and 354.1 s could be reproduced with $l$=0 ($k$=1) and $l$=1 
($k$=1) respectively.
However this solution was not unique due to the small number of detected 
frequencies and other solutions could not be excluded.

\vspace{13.6mm}

\section{Observations, reduction and analysis}

\begin{figure*}[th]
\vspace{12.6cm}
\includegraphics{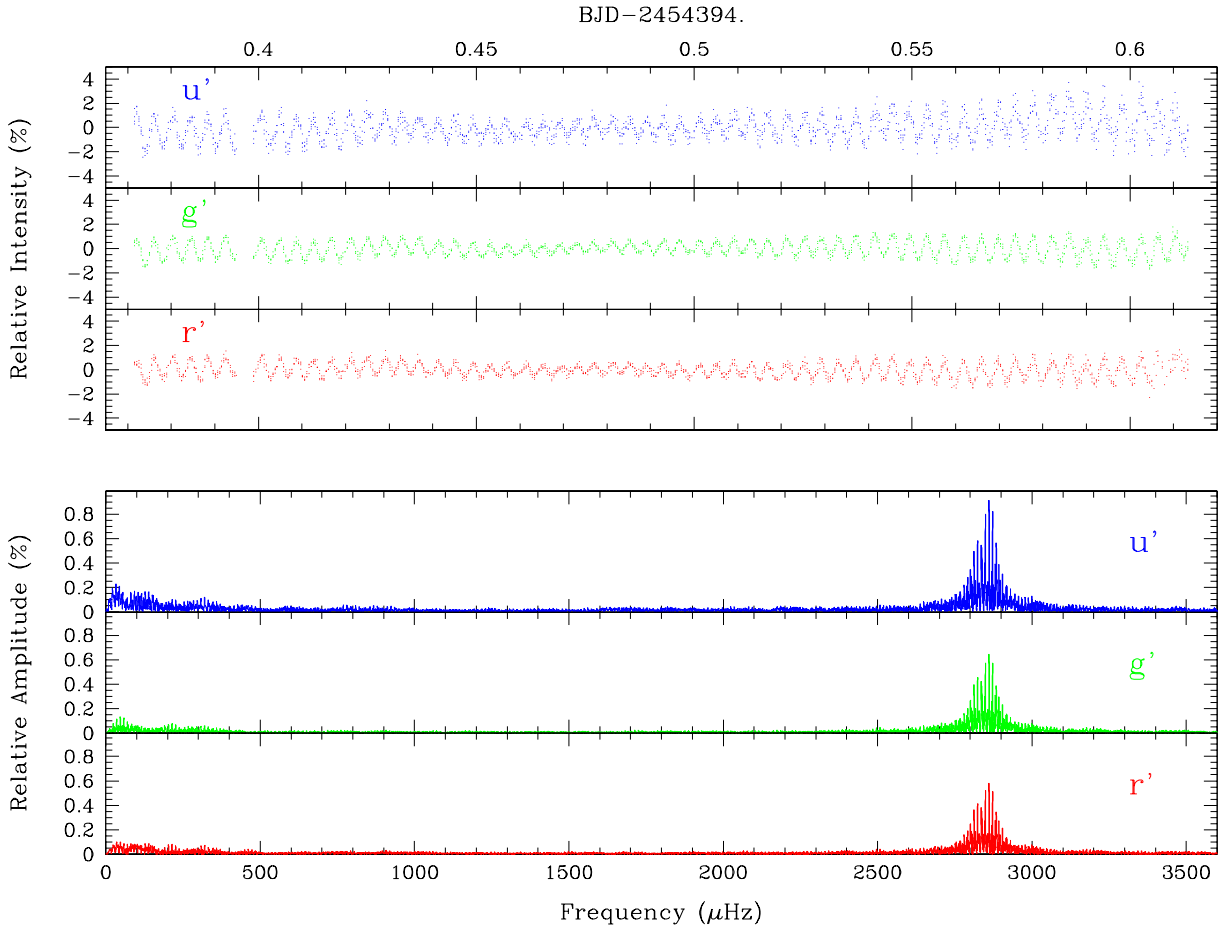}
\caption{Upper panels: representative u'g'r' light curves (20 October 2007), 
with beating effects clearly visible.
Lower panels: amplitude spectra of the whole 8-nights run showing the two 
regions of excited $p$- and $g$-modes.}
\label{f1}
\end{figure*}

V391~Peg was observed for 8 consecutive nights, from October 16 to 23, 2007,
using ULTRACAM (Dhillon et al. 2007) at the 4.2 m William Herschel telescope 
(WHT).
 In total we collected 260,592\, exposures in three
photometric bands ({\it u'g'r'}) of the SLOAN system, with exposure times 
between 1.2 and 1.6~s and a dead time of only 25~ms between one frame and the 
next.
The reduction was performed using the ULTRACAM pipeline (see, for example, 
Littlefair et al. 2008), including bias and flat field correction and aperture
photometry.
Then we computed differential photometry, dividing the target's counts
by the counts of a comparison star, we binned the data to an effective
exposure time of 10 s and we performed a correction for the residual 
extinction.
The last step is crucial for the $g$-modes which have periods of $\sim$0.5-2 h
and are particularly disturbed by variations that occur on similar
time scales.
Finally we applied the barycentric correction to the times.
More details regarding observations and data reduction will be given in a
forthcoming paper (Silvotti et al. in preparation).
At the end of the process, for each filter we obtained a single file with 2 
columns: barycentric julian date and fractional brightness variation.

\begin{figure*}[th]
\vspace{11.0cm}
\includegraphics{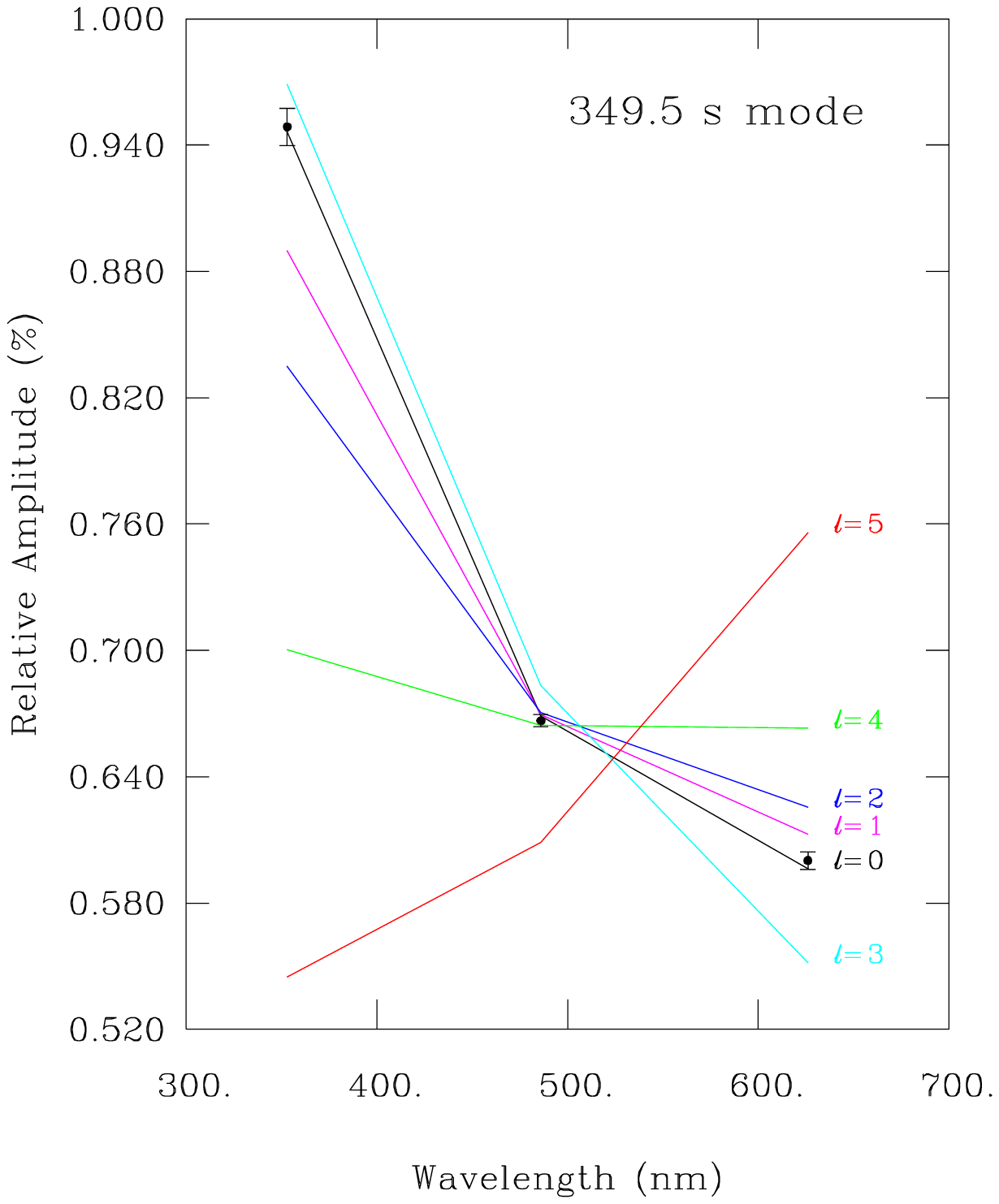}
\includegraphics{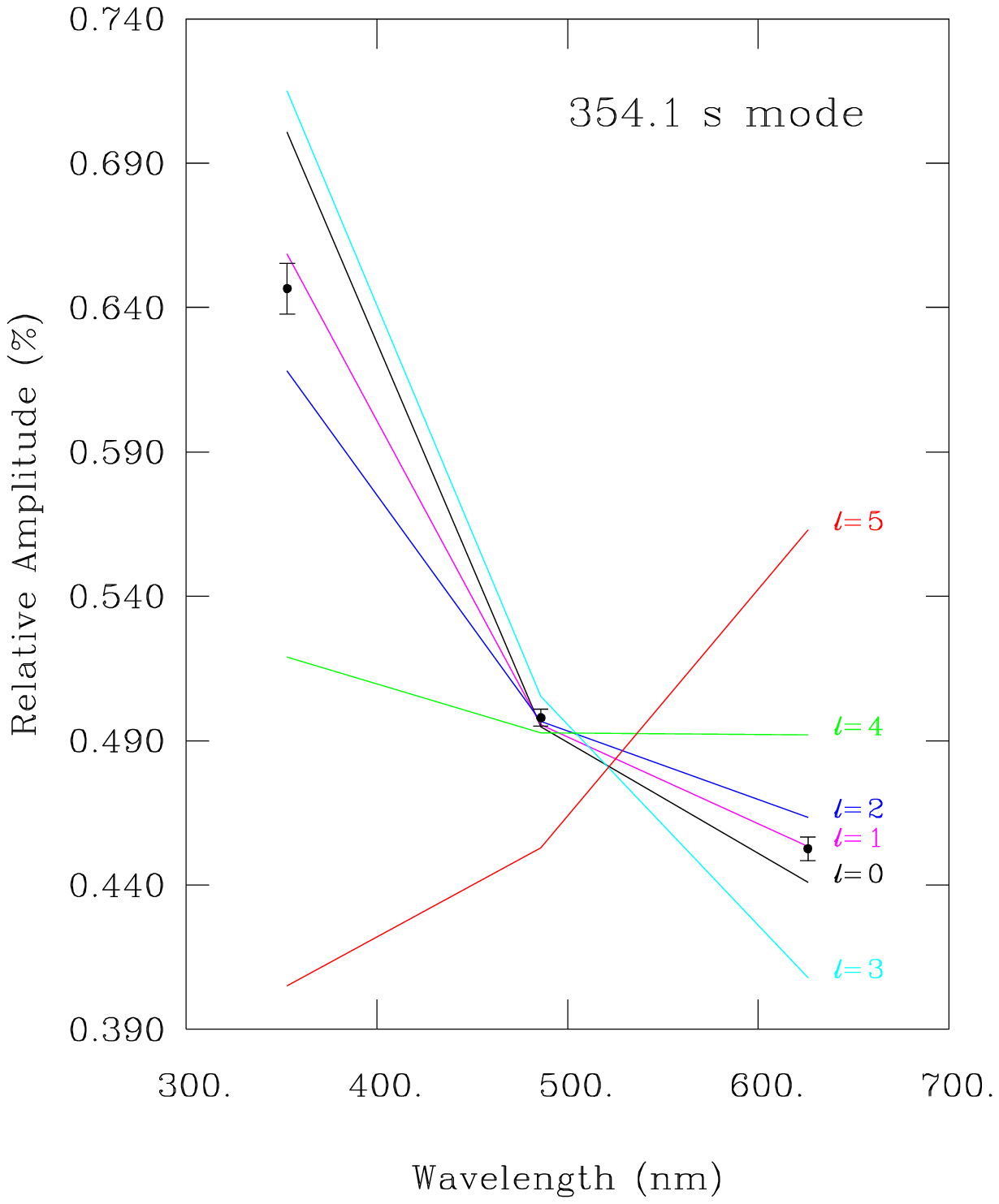}
\caption{Fit to the two dominant modes of V391~Peg, including non-adiabatic 
effects.}
\label{f2}
\end{figure*}

These files were analyzed using Period04 (Lenz and Breger 2005) in order to 
determine the pulsation frequencies and the amplitudes in the different bands.
A portion of the light curves and the amplitude spectra in the three 
photometric bands are shown in Fig.~1.
The frequencies obtained were compared with those obtained from previous runs
and indeed we found a perfect agreement except for f4 and f5: the value  
2921.816 $\mu$Hz found previously for f4 (Silvotti et al. 2002) is now
2910.272 $\mu$Hz, indicating that the old value was a 1-day alias of the
correct frequency.
The new value of f4 is confirmed by two other independent runs with high
frequency resolution at the 3.6 m TNG (August 2008) and 1.3 m MDM (October 
2007) telescopes (Silvotti et al. in preparation).
F5 (2738.015 $\mu$Hz, Silvotti et al. 2002) is not found in any of the new 
observations and the WHT/ULTRACAM data suggest 2 new low-amplitude 
frequencies.
An updated list of frequencies, including also the low-frequency $g$-modes, 
will be published (Silvotti et al. in preparation).

Using the improved list of frequencies, we measured the amplitudes of the
various frequencies by means of least-square sinusoidal fits.
In this paper we concentrate only on f1 and f2, for which the errors on the 
amplitudes are sufficiently small to obtain significant results.

\section{Comparison with theoretical amplitudes and mode identification}

The amplitudes obtained have been compared with theoretical amplitudes 
calculated following the same procedure as described in Randall et al. (2005).
As input we took the atmospheric parameters from {\O}stensen et al. 2001
(\teff=29,300 K, \logg=5.4).
With the very small error bars of the ULTRACAM amplitudes, the quality of the
fit is very sensitive to the exact input values of the atmospheric parameters.
However, taking into account the \teff\ \logg\ uncertainties, we never obtain 
any change in the mode identification.
The same is true when we use slitghtly different atmospheric parameters 
obtained by one of us (G.F.) on the basis of spectra acquired at the 
90 inch telescope at Kitt Peak (\teff=30,000 K, \logg=5.46)\, or\, at\, the \,MMT\,
(\teff=29,900 K, \logg=5.51), made available to us by Betsy Green (private 
communication).

The monochromatic\, atmospheric\, quantities were then convolved over the ULTRACAM
filters, taking into account the filter response curves as well as the quantum
efficency of the CCDs, and the transparency curve for a representative
observatory site (we used Cerro Tololo which is at a similar altitude as the 
La Palma observatory).

Non-adiabatic effects were computed as they significantly influence the
theoretical colour-amplitudes.
They were estimated using the equations given in Randall et al. (2005) from 
the adiabatic and non-adiabatic eigenfunctions calculated from second 
generation envelope models (Charpinet et al. 1997), using the following 
stellar parameters: 
\teff=29,300 K, \logg=5.4, total stellar mass M$_{\ast}$=0.48 \msun\ and
fractional thickness of hydrogen-rich envelope $\log q(H)$=$-2.95$, although 
the two last values do not really influence the results much, as long as they 
take on reasonable values.
The model was then analysed with adiabatic and non-adiabatic pulsation codes,
and the non-adiabatic quantities {\it R} and $\Psi_{\it T}$ (defined in 
Randall et al. 2005) were computed for the period spectrum of the model.
Since {\it R} and $\Psi_{\it T}$ are not dependent on the degree index while 
they depend quite strongly on the period of the mode in question, their value 
were interpolated in period space to find their optimal values for the 
observed periods.

The wavelength-integrated atmospheric quantities and the non-adiabatic
parameters were then used to calculate the theoretical amplitudes.
As a last step, the theoretical amplitudes were fit to those observed using 
the $\chi^2$ minimisation technique described in Randall et al. (2005).

The results, shown in Fig.~2, indicate a unique solution for the two main 
pulsation modes of V391~Peg.
The 349.5~s period is a radial mode:
$\chi^2$($l=0$)=1.5, $\chi^2$($l=1$)=53.6, $\chi^2$($2 \leq l \leq 5$)$>$170.
The 354.1~s period is a dipole mode and again there is only one solution
compatible with the data: 
$\chi^2$($l=1$)=2.5, $\chi^2$($l=2$)=17.1, $\chi^2$($l=0$)=47.4, 
$\chi^2$($3 \leq l \leq 5$)$>$180.
These numbers translate into a value of the quality-of-fit parameter
(Press et al. 1986) Q $\ll$ 0.001 for both modes when we use an $l$ value 
different from 0 and 1 respectively.

%\vspace{2mm}
\section{Discussion}

Thanks to the high quality of the data, this is the first time that the mode 
degree index has been uniquely identified from multicolor photometry for the 
two main modes of a star as faint as V391~Peg (V=14.6).
%
%There have been several attempts to determine the degree index from 
% multicolor photometry (e.g. Koen 1998; Jeffery et al. 2004, 2005), but 
%
To our knowledge, conclusive results were obtained only for two brighter 
stars:
KPD~2109+4401 (V=13.4, Randall et al. 2005, see also Jeffery et 
al. 2004) and Balloon 090100001, the brightest known sdBV star with V=12.1 
(Baran et al. 2008, Charpinet et al. 2008).

The results reported in this article confirm that multicolor photometry can 
set useful identification constraints on the pulsation modes of sdB rapid 
pulsators, provided that the data have a high enough quality.
ULTRACAM on a 4~m class (or larger) telescope is an ideal instrument for such 
studies.

\acknowledgements

R.S.\, thanks\, Stuart Littlefair for his help in data reduction.
R.S. acknowledges support from HELAS for participating to this conference.

\newpage%%%%%%%%%%%%%%%%%%%%%%%%%%%%%%%%%%%%%%%%%%%%%%%%%%%%%%

%\appendix
%
%\section{}

\end{document}